\documentclass[]{spie}  

\usepackage{amsmath,amsfonts,amssymb}
\usepackage{graphicx}
\usepackage[colorlinks=true, allcolors=blue]{hyperref}
\usepackage{astro_bib_macro}

\usepackage{booktabs}

\usepackage{listings}

\usepackage{color}

\title{Programmed but Arbitrary Control Minimization of Amplitude and phase for speckle Nulling (PACMAN)}

\author{Iva Laginja\supit{a}, Raphaël Pourcelot\supit{b}
\skiplinehalf
\supit{a} LESIA, Observatoire de Paris, Universit\'{e} PSL, Sorbonne Universit\'{e}, Universit\'{e} Paris Cit\'{e}, CNRS, 5 place Jules Janssen, 92195 Meudon, France\\
\supit{b} Space Telescope Science Institute, 3700 San Martin Drive, Baltimore, USA
}

\authorinfo{Further author information, send correspondence to Iva Laginja: E-mail: iva.laginja@obspm.fr}

\begin{document} 
\maketitle

\begin{abstract}
We revive a cross-platform focal-plane wavefront sensing and control algorithm originally released in 1980 and show that it can provide significant contrast improvements over conventional control methods on coronagraphic instruments. Its simplicity makes it applicable to various coronagraph models and we demonstrate its working on a classical Lyot coronagraph and a phase-apodized pupil Lyot coronagraph, both in simulation and in laboratory experiments. Surprisingly, it had been forgotten for decades, but we present its unbeatable advantages considering the increase in computational power in the last 40 years. We consider it a major game changer in the planning for future, space-based high-contrast imaging missions and recommend it be intensively revisited by all readers\cite{Google2010}.
\end{abstract}


\section{Introduction}
\label{sec:introduction}

The search for exoEarths and atmospheric biomarkers is an exciting field in astronomy, but capturing photons from planets near their host stars requires significant improvements in imaging capabilities. Instruments must achieve contrast levels of at least $10^{-10}$ at a separation of $\sim0.1$ arcsec or less from the star, which will require telescopes with large collecting areas and segmented primary mirrors. Coronagraphs\cite{Vilas1987CoronagraphAstronomicalImaging,Lyot1939StudySolarCorona} are the favored method to achieve these high-contrast levels, but they are sensitive to residual wavefront aberrations that can generate speckles of light in the imaging focal plane. These can be misinterpreted for exoplanetary signals and pose a very significant source of noise in high-contrast observations. To create a zone of deep contrast in the final image, a ``dark hole'' (DH), coronagraphy must be combined with active wavefront sensing and control (WFS\&C\cite{Mazoyer2018a:ActiveCorrectionApertureI,Mazoyer2018b:ActiveCorrectionApertureII,Groff2016MethodsLimitationsFocal}). The Habitable Worlds Observatory (HabWorlds) has been selected by the NASA Astro2020 Decadal Survey\cite{astro2020} as the mission concept to achieve these goals from space, while the Nancy Grace Roman Space Telescope (RST\cite{Krist2015OverviewWFIRSTAFTA}) is working toward shorter-term demonstrations at more moderate contrast levels on the order of $10^{-8}$--$10^{-9}$. The next generation telescopes on the ground, the extremely large telescopes, such as the Extremely Large Telescope (ELT\cite{2018CirasuoloELT1, 2019CayrelELT2, 2019VernetELT3}), the Thirty-Meter Telescope (TMT\cite{Wolff2019TMT}) or the Giant Magellan Telescope (GMT\cite{McCarthy2016GMT1, Fanson2018GMT2, Fanson2020GMT3}) are pursuing similar goals albeit at more modest contrast levels compared to their space-based counterparts, as they will be limited by the influence of the Earth's atmosphere.

Achieving extreme contrast levels requires excellent stability against wavefront errors over a range of temporal and spatial frequencies. To enable the creation of a DH, as well as to stabilize the contrast once the DH has been established, many different control techniques have emerged. They usually rely on a single or a pair of deformable mirrors (DMs) to modulate the wavefront phase and/or amplitude based on an estimation of the aberration to correct\cite{Beaulieu2020HighContrastSmall,Will2021JacobianfreeCoronagraphicWavefront}. The earliest forms of wavefront control (WFC) techniques were simple both in performance as well as application as they only required a measurement of the focal plane intensity to calculate correction steps for the DM\cite{Malbet1995HighDynamicRangeImagingUsing}. Collectively called ``speckle nulling'', part of the simplicity of this method lies in the fact that it uses linearized equations, thus discarding higher-order terms in the development of the electric field with aberrations, and unlike most methods used today, it does not rely on a model. It is thus simple to implement, and it has been demonstrated on several high-contrast testbeds\cite{Soummer2018HighcontrastImagerComplex,Leboulleux2017ComparisonWavefrontControl,Belikov2006Toward1010Contrast,Trauger2004CoronagraphContrastDemonstrations}.

Today, more advanced strategies exploit an estimation of the electric field in the focal plane to determine the control commands for the DMs. The most common algorithms include electric field conjugation (EFC\cite{Give'on2007ClosedLoopDM}), where the DM correction commands is calculated under the objective to minimize the total energy across all pixels in the designated DH area. Similarly, stroke minimization (SM\cite{Pueyo2009OptimalDarkHole}) seeks to improve the DH contrast in the same way, but it does so by also minimizing the total stroke of the DM actuators.

However, one precursor method, way ahead of its time, has been historically overlooked for its potential to reduce the speckle noise in coronagraphic images. In this paper, we present the Programmed but Arbitrary
Control Minimization of Amplitude and phase for speckle Nulling (PACMAN), an algorithm designed to eliminate the speckles in the coronagraphic field of view. PACMAN is based on a lightweight cross-platform algorithm, and is autonomous in the sense it only requires a data-based calibration to remove polluting speckles. We introduce this method in Sec.~\ref{sec:killing-speckles-with-pacman}, show results in simulation and hardware in Sec.~\ref{sec:results}, discuss its strengths and weaknesses in Sec.~\ref{sec:real-data} and summarize our work in Sec.~\ref{sec:summary}.

\section{Killing speckles with Pacman}
\label{sec:killing-speckles-with-pacman}

\subsection{Wavefront control using a forward model of the optical system}
\label{subsec:HCI-instrument}

One of the main problems for any control strategy is to determine the adequate command $\mathbf{u}_k$ to be sent to the DMs to obtain the relevant change in the focal plane (i.e., improve contrast). Constructing a well-defined forward model by using Fraunhofer formalism, or Fresnel formalism in the case of out-of-pupil DMs, a single intensity measurement in the focal plane yields an indetermination on the exact state of the wavefront. This is thus insufficient to conduct WFC. One way to tackle the problem is to build a forward linear problem, for example between the actuators of DMs and the electric field in the focal plane, under the form of a matrix $J$. This is often called ``interaction matrix'' or ``Jacobian matrix''. Inverting $J$ provides the control matrix $K$, that is a linear backward model enabling the computation of the DM commands. In the following, we briefly describe the forward model used to make PACMAN run in the focal plane and the associated Jacobian and control matrices.

\begin{figure}
    \centering
    \includegraphics[width = 0.8\textwidth]{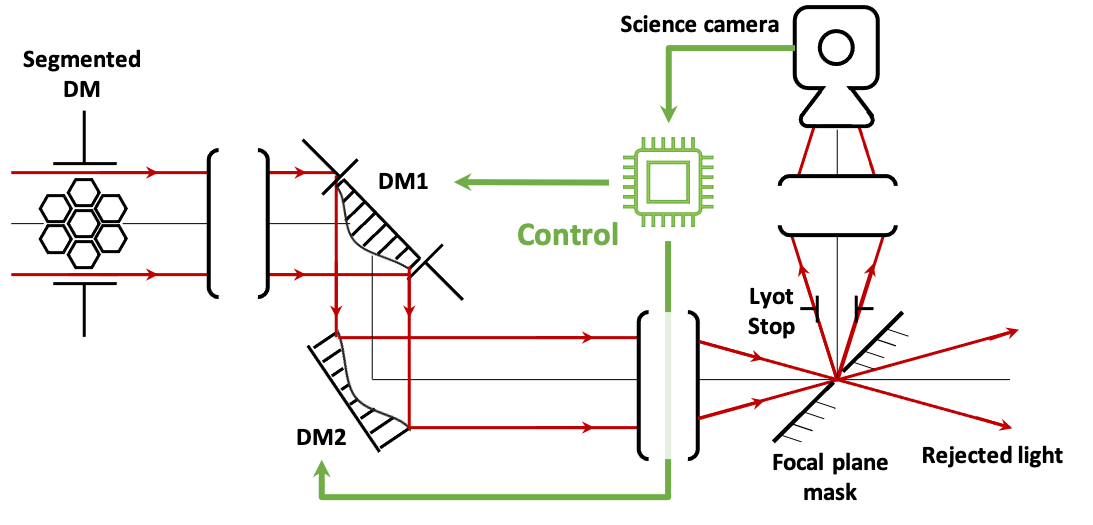}
    \caption{Generalized design of a HCI instrument. Light enters the instrument through an aperture on the left and is modulated by two DMs, as well as coronagraphic masks located both in focal planes as well as in pupil planes. The sequence of these optics does not matter, as PACMAN has been shown to be able to gobble up all flavors of speckles.} 
    \label{fig:hci_layout}   
\end{figure}

A generic high-contrast imaging (HCI) instrument consists of a combination of coronagraphic masks and a pair of DMs as shown in Fig.~\ref{fig:hci_layout}.
To allow for an easier analytic treatment, we assume one of the DMs to be located in a pupil plane, although this is not a constraint in practice. Propagation between pupil planes and focal planes are treated with Fourier transforms representing Fraunhofer (far-field) diffraction, while near-field propagation to optics outside of these planes (like the out-of-pupil DM) are treated with an angular spectrum operator. The coronagraph is a linear operator $\mathcal{C}$ and the exact location of its masks with respect to the DMs does not matter, as long as they are constrained to focal and pupil planes of the optical system.

Following the development of of Will et al.\cite{Will2021JacobianfreeCoronagraphicWavefront}, we approximate the electric field in the pupil plane at a control iteration $k$, $E(\mathbf{p})$, where $\mathbf{r}$ is the pupil-plane coordinate, under a small-aberration assumption to:
    \begin{equation}
    E_k(\mathbf{r}) = A(\mathbf{r})\ e^{i\phi_{k-1}(\mathbf{r})} [1 + g(\mathbf{r})] + i A(\mathbf{r})\ e^{i\phi_{k-1}(\mathbf{r})}\ \delta \phi_k(\mathbf{r}),
    \label{eq:isolated-pup-field}
    \end{equation}
where $A(\mathbf{r})$ is the transmissive pupil function, $g(\mathbf{r}) = \alpha + i \beta$ its amplitude and phase aberrations, $\phi_k(\mathbf{r})$ the DM phase from a previous iteration $k-1$ and $\delta \phi_k(\mathbf{r})$ a small change of the DM phase in the current iteration. We propagate Eq.~\ref{eq:isolated-pup-field} to the image plane by applying the coronagraph operator $\mathcal{C}$:
    \begin{equation}
    E_{k}(\mathbf{s}) = E_{ab, k}(\mathbf{s}) + \ i \mathcal{C}\{A(\mathbf{r})\ e^{i\phi_{k-1}(\mathbf{r})}\ \delta \phi_k(\mathbf{r})\},
    \label{eq:e-field-image-plane-no-actuators}
    \end{equation}
where $\mathbf{s}$ is the image-plane coordinate, $E_{ab, k}(\mathbf{s})$ the propagated aberrated pupil function and here dubbed aberrated electric field. Assuming linearity within a single DM actuator as well as across all actuators on a single DM, we can express the differential DM phase through a linear combination of the DM influence functions $f_q(\mathbf{r})$ of all actuators $N_{act}$, indexed by $q$, and with actuator amplitude $a_q$:
    \begin{equation}
    \delta \phi = \frac{2\pi}{\lambda} \sum_{q=1}^{N_{act}} \delta a_q f_q(\mathbf{r}).
    \label{eq:differential-phase-from-actuators}
    \end{equation}
We can thus substitute Eq.~\ref{eq:differential-phase-from-actuators} into Eq.~\ref{eq:e-field-image-plane-no-actuators} to obtain the full electric field in the focal plane:
    \begin{equation}
    E_{k}(\mathbf{s}) = E_{ab, k}(\mathbf{s}) + \frac{2\pi i}{\lambda}\ \sum_{q=1}^{N_{act}} \delta a_q \mathcal{C}\{A(\mathbf{r})\ e^{i\phi_{k-1}(\mathbf{r})}\  f_q(\mathbf{r})\}.
    \label{eq:e-field-image-plane-with-dm}
    \end{equation}
Discretizing the electric field and associated parameters onto a camera with $n$ discrete pixels, indexed by $p$, we can write Eq.~\ref{eq:e-field-image-plane-with-dm} in vectorial form:
    \begin{equation}
    \mathbf{e}_{im, k} = \mathbf{e}_{ab, k} + J_{IN, k}\ \mathbf{u}_k,
    \label{eq:image-efield-matrix}
    \end{equation}
where $\mathbf{e}_{im, k}$ is the discrete electric field vector in the focal plane at iteration $k$, $\mathbf{e}_{ab, k}$ the discretized aberrated field, $J_{IN, k}$ is the Jacobian matrix for the in-pupil DM, and $\mathbf{u}_k$ the vector of actuator commands.
Following the development in Will et al.,\cite{Will2021JacobianfreeCoronagraphicWavefront} we extend this model with the second DM by using an augmented Jacobian matrix, $J_k$, that concatenates the matrices from each DM:
    \begin{equation}
    J_k = [J_{IN, k}~~J_{OUT, k}]
    \label{eq:augmented-jacobian}
    \end{equation}

Once the Jacobian has been measured, usually either through a forward finite difference (more common for space PACMAN) or a central difference approximation (more common for ground-based PACMAN) it can be inverted to obtain the control matrix $K$. However, in the common case where the number of measurements, here pixels in the focal plane, is larger than the number of actuators, inverting $J$ is an ill-conditioned problem, and requires a single value decomposition (SVD) to identify the most prominent sensed modes (the singular modes). To avoid instabilities in the control, less-sensed modes then need to be filtered out, either by truncating the pseudoinverse, or by using for example a Tikhonov regularization.

\subsection{The PACMAN algorithm}
\label{subsec:pacman-algorithm}

The goal of an HCI instrument as described in Sec.~\ref{subsec:HCI-instrument} is to attenuate the on-axis star light while retaining any physical off-axis signal in the final focal plane of the instrument, aided by a WFS\&C system.
Speckle management in the focal plane has been made possible with the advent of mainstream computers. Before nulling, pioneer strategies have been developed in the early 1970s to move them around the focal plane, with one of the earliest being ``Pong'' \cite{Pong1972, Diallo2017Pong}. After this initial success, and the steady increase in computing power, more advanced algorithms have been developed, with for example PACMAN, first presented in the early 1980s \cite{Namco1980,Goodman1983PacMania}. PACMAN has since experienced a world-wide success, being ported to many hardware and experiment infrastructures like Atari\cite{montfort2020racingAtari, culler1983copyrightAtari} and Nintendo\cite{NintandDoe1990}, and being extensively studied \cite{Rohlfshagen2018PacmanConquers}. Overall, more than 50 versions have been published, with more than 50 million versions sold to principle investigators all around the world. Even if the original code was proprietary, finding open-source equivalents for this influential and iconic masterpiece is now widely possible.

To deploy PACMAN and have it eat up dots of light, we iterate over all identifies speckles to remove the light from the focal plane. The flow of the loop is displayed in Fig.~\ref{fig:hci_layout}. At an iteration $k$, after identifying the brightest speckle in the focal plane image, described by the vector of pixels $\mathbf{x_k}$, we compute a transformation $T_k$, dependent on the speckle position, such that the new image yields the suppression of the bright speckle:
    \begin{equation}
    \mathbf{x_{k+1}} = T(\mathbf{x_k}).
    \label{eq:speckle-suppresion}
    \end{equation}
We chose $T_k$ as a succession of focal plane modifications, $P_i$, determined as optimal by the literature. To obtain the corresponding commands $\mathbf{c_i}$ to be applied to generate the $P_i$ modification, we use the back-propagation:
    \begin{equation}
    \mathbf{c_i} = K_k \cdot P_i,
    \label{eq:back-propagation}
    \end{equation}
where $\cdot$ denotes the matrix product. We note how $K$ is the control matrix, calculated from the inverse of the augmented Jacobian, $J_k$, as detailed in Sec.~\ref{subsec:HCI-instrument}.

Finding the speckle we want to remove at iteration $k$ can be a challenging task. While the first ones are usually easy to spot, the dimmer ones are more difficult to identify, especially if background or incoherent light is polluting the image. In particular, operating PACMAN in a dark environment, windows closed, is usually recommended even if this can induce an addiction to this algorithm.


Once the spot to remove is identified, we project its location on a focal-plane grid called ``maze'' that we define in terms of diffraction units $\lambda / D$, with $\lambda$ the wavelength and $D$ the entrance pupil diameter. The maze is way too often limited by an outer wall, the ``Nyquist wall'', that defines the DH, and that prevents PACMAN from moving outwards. Moving beyond these walls usually results in scattering of light in multiple locations, which is counterproductive. Similarly, moving the wall itself to increase the maze size is difficult, since it requires very important hardware investments, in term of hardware, manpower for the setup, and most importantly, cabling solutions.

In this context, solving the route for PACMAN is straightforward using a Djikstra \cite{dijksta1959note} algorithm. In the case of the necessity to extract multiple sources, one must be more careful since this relates to a traveling salesman problem that can become extremely expensive computationally. However, external constraints can happen even once the shortest path has been found. These control instabilities, usually called ``ghosts'', tend to get closer to PACMAN as time evolves, slowly but surely threatening the relevance of the algorithm. While ghosts can be a cumbersome problem in most optical applications, PACMAN actually identifies four particular flavors of them that each have a different performances on the light attenuation efficiency: Blinky, Pinky, Inky and Clyde. They differ in their particular strategy of cornering PACMAN and making the algorithm stall. While PACMAN is operable in broadband light of up to 50\% in visible wavelengths, there is an inherent chromaticity associated with the different ghosts. While Blinky obstructs PACMAN at wavelengths around 700~nm, Pinky only becomes a problem in dual combinations of wavelengths around 700~nm and 400~nm. Inky inflicts issues at a wavelength around 500~nm and Clyde around 600~nm.

Three ways are possible to mitigate the ghosts: (i) By constantly moving PACMAN from one speckle to the other, one prevents the ghosts from interfering; (ii) by solving the control matrix propagation for paths that avoid them on purpose, even if this can get computationally expensive; (iii) by ingesting power pellets, which are similar to speckles but that cannot be removed with this nulling technique, that we call ``planets''.

Overall, the combination of maze structure, ghost behavior, limited resources, and power pellets creates a challenging environment for PACMAN to navigate, making the algorithm both programmed as well as arbitrary.


\section{High scores in simulation and on hardware}
\label{sec:results}

In this section, we present our results obtained with PACMAN on the High-contrast imager for Complex Aperture Telescopes (HiCAT\cite{Soummer2022HighContrastImager}) testbed, hosted at the Space Telescope Science Institute in Baltimore, USA. We show both simulated as well as experimental data. HiCAT now supports several types of Lyot-style coronagraphs, all working in conjunction with a segmented deformable mirror of type ``IrisAO'', used to emulate a segmented primary mirror. In total, we present results without a coronagraph, with a classical Lyot coronagraph (CLC) that uses a circular, hard-edge focal-plane mask (FPM), and a phase-apodized pupil Lyot coronagraph (PAPLC\cite{Por2020PhaseApodizedPupil}), using a knife-edge FPM. \textbf{Associated video files showing the full PACMAN loops of all four data sets can be accessed on YouTube}\footnote{\url{https://www.youtube.com/playlist?list=PLBmfHGsq5CfciJFPrtc_BCzWCLslXdrno}} (also accessible with the QR code in Fig.~\ref{fig:image_Sequence}), or downloaded from Zenodo\cite{PACMAN2023Zenodo}.

\subsection{Simulation results}
\label{subsec:simulation-results}

Since the CLC is the simplest possible coronagraphic mode on HiCAT, we used it for our first simulated benchmark test. We then also simulated PACMAN routes for the PAPLC, to prepare to go on hardware. The HiCAT optical model is written in a user-friendly, modular setup using the hcipy Python package\cite{Por2018HighContrastImaging}.

\begin{figure}
    \centering
    \includegraphics[width=\linewidth]{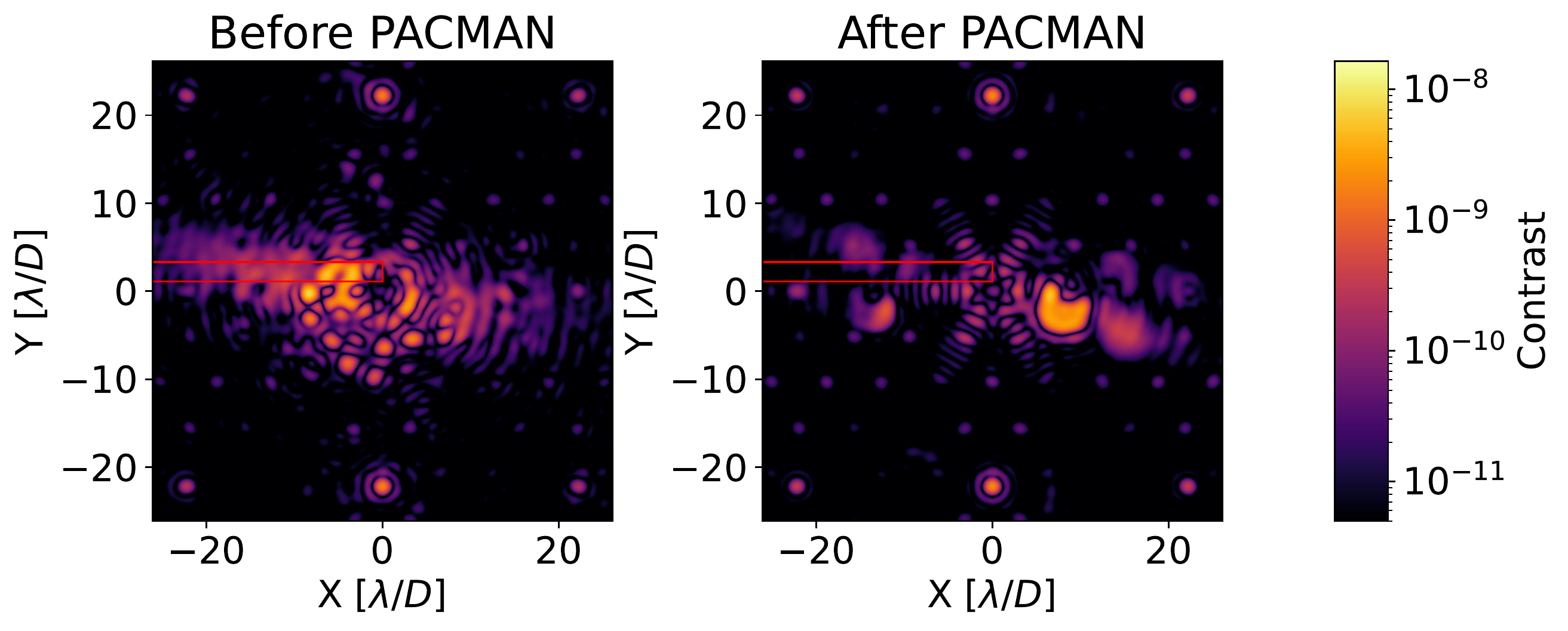}
    \caption{Simulated images of typical focal-plane speckles polluting the image of an observed star in the case of a CLC coronagraph. The ares marked by rectangular shapes were used to calculate the radial profiles in Fig.~\ref{fig:clc_cut}. \textit{Left:} Pre-PACMAN image. \textit{Right:} Post-PACMAN image. \textbf{See online video material for animated videos of the PACMAN loop\cite{PACMAN2023Zenodo}}.}
    \label{fig:sim_clc_initial_img}
    \end{figure}

Examples of CLC focal-plane images polluted with speckles are shown in Fig.~\ref{fig:sim_clc_initial_img}.
We can identify several kinds of speckles. The eight outer ones by the edge of the frame, arranged on a square grid, are diffraction spots originating from the continuous DMs. These one are located on the outside of the walls of the maze and can thus not be destroyed by PACMAN. The less luminous points, arranged on an hexagonal grid in this image come from the segmented DM, and lie within the range of PACMAN. However, we do not have rock-solid results on these speckles yet, even if the gobble machine we released onto this multi-million dollar project looks very promising. Finally, the group of most luminous speckles close to the center of the image usually arise due to instabilities in the instrument, like the loose hands of some postdocs while operating the DMs, and they are the primary target for PACMAN. 

\begin{figure}
    \centering
    \includegraphics[width=.5\linewidth]{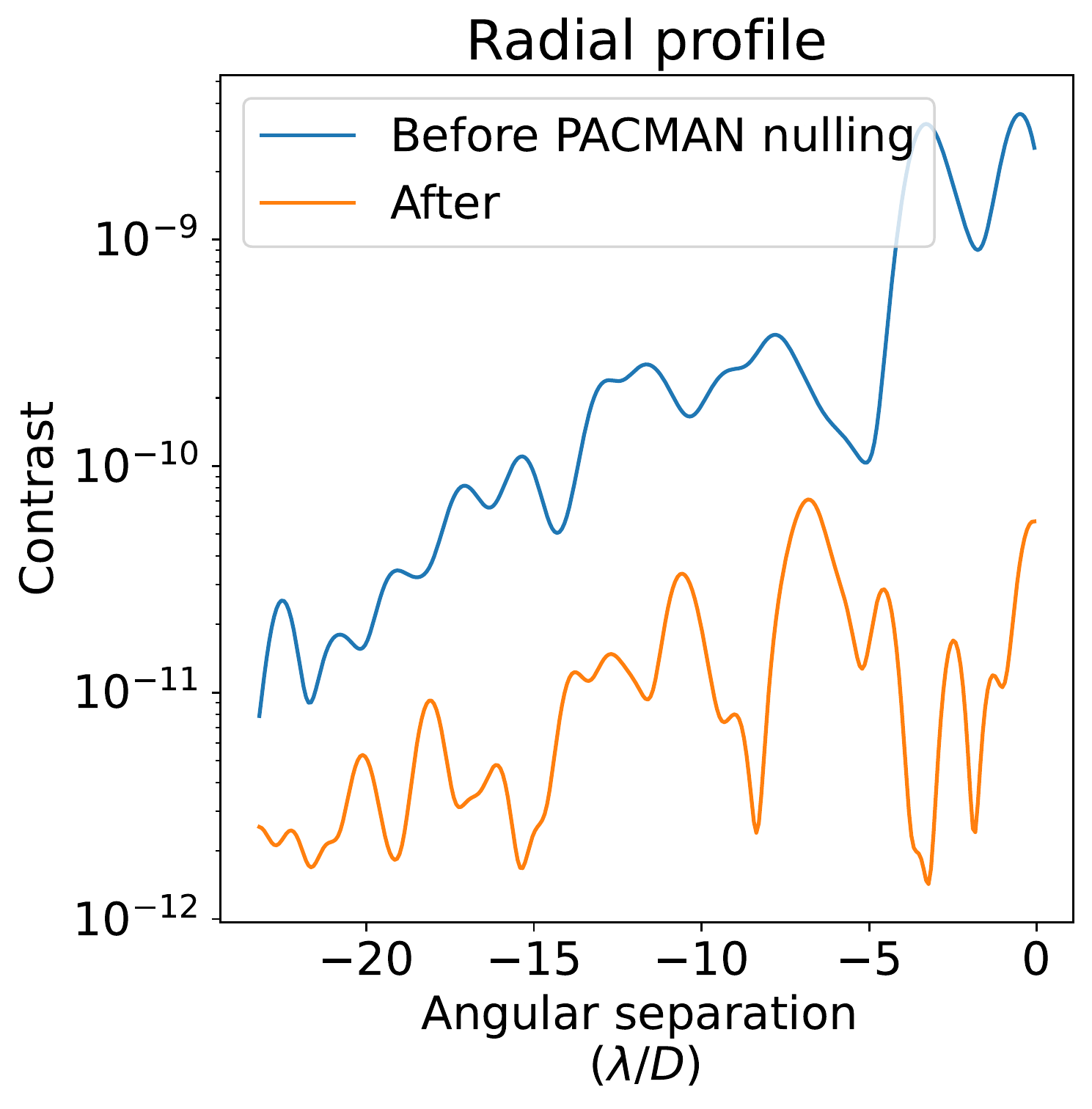}
    \caption{Radial contrast plots of simulated CLC images before and after PACMAN ran on it. We observe a significant improvement in contrast at all angular separations.}
    \label{fig:clc_cut}
    \end{figure}

Running PACMAN on this initial speckle distribution proves very effective, as is visible in Fig.~\ref{fig:clc_cut}. It provides a contrast improvement by a factor of up to three orders of magnitude. A radial cut of the initial and final image is presented in Fig.~\ref{fig:clc_cut}, showing a clear bump in the intensity where the speckles are at separations closer than 4 $\lambda/D$, and are completely removed after the passage of PACMAN.
A comparison of the before and after curves shows the astounding gains achievable with this method. Upon closer inspection of the ``after'' image (Fig.~\ref{fig:sim_clc_initial_img}, right), we can easily identify some of the ghosts sneakily going after PACMAN, but a smart choice of its route enables a significant improvement in contrast.

\begin{figure}
    \centering
    \includegraphics[width=\linewidth]{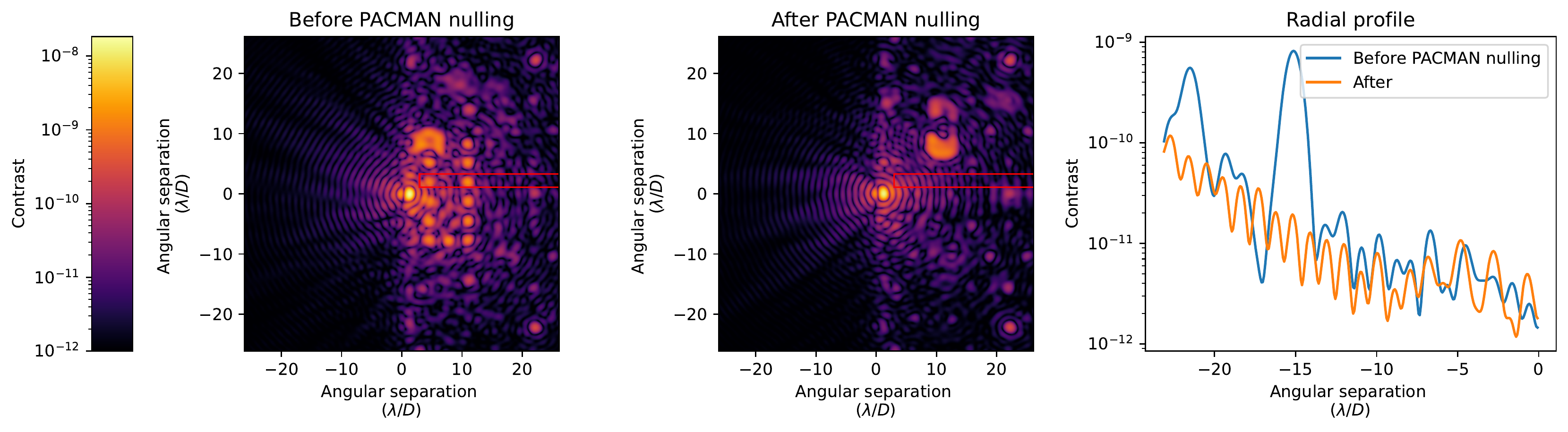}
    \caption{Simulated PACMAN run with a PAPLC. Due to the nature of the coronagraph, an additional wall has been introduced, limiting PACMAN to one half of the focal plane. \textit{Left:} First iteration of PACMAN, no speckles have been suppressed so far. \textit{Middle:} After a run of PACMAN. \textit{Right:} Plot of the red rectangle averaged over the vertical dimension.}
    \label{fig:sim_paplc_cut}
\end{figure}

Since the current coronagraph installed on HiCAT is a PAPLC\cite{Por2020PhaseApodizedPupil}, we proceeded with a simulation of PACMAN with DM-induced phase apodization, and a knife-edge FPM. The before and after images, as well as the associated radial profiles are shown in Fig.~\ref{fig:sim_paplc_cut}.
The use of a knife-edge FPM, there is an additional wall introduced to the maze on top of the Nyquist wall, as PACMAN is now constrained to one side of the focal plane only. While this introduces a relaxation in the path projection, it also limits the area potentially inhabited by planetary power pellets.

The residual speckles, present in the simulated data of both the CLC as well as the PAPLC, are due to the segmentation, which PACMAN still needs to learn to digest. However, since this vintage algorithm is able to provide us with an average contrast level of $10^{-10}$ in simulations already, we lean back, progress through the levels and see how the problem gets more and more complex as we deploy PACMAN on hardware.

\subsection{Hardware results}
\label{subsec:hardware-results}

\begin{figure}
    \centering
    \includegraphics[width=\linewidth]{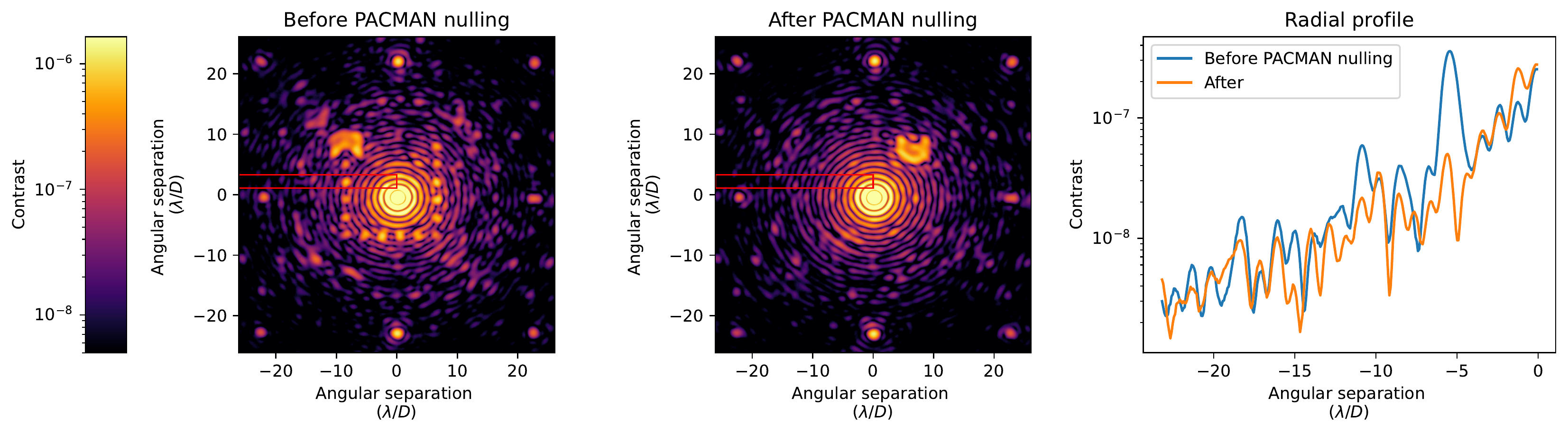}
    \caption{Experimental PACMAN run without an FPM. Even though this provides a worse contrast overall, this configuration enables the cleaning of a larger field of view and can be adapted to many imaging configurations. \textit{Left}: First iteration of PACMAN, no speckles have been suppressed so far. \textit{Middle}: After a run of PACMAN. \textit{Right}: Plot of the red rectangle averaged over the vertical dimension. We se a distinct improvement of the contrast at a distance of 5 $\lambda/D$.}
    \label{fig:direct_cut}
    \end{figure}

Using the HiCAT facilities, we were able to experimentally test our approach. The seamless transition of simulations to experimental implementation was enabled by the dual-mode operations setup with a seamless toggle between the optical model and the true hardware testbed. To make sure PACMAN is behaving properly without undesirable effects due to a coronagraphic FPM, we first used it in a non-coronagraphic configuration, without an FPM in the beam. The before and after images, as well as a radial profile through the image, are shown in Fig.~\ref{fig:direct_cut}.
While the contrast levels are not as good as with a FPM, around $10^{-7}$ at shorter inner working angles, the action of PACMAN is easily noticeable, with an obvious improvement in the diffraction pattern. A horizontal cut, created by averaging the contrast along the vertical axis in a section marked by a red rectangle in the images, shows an impressive attenuation of the bright speckle at $5\lambda/D$ by a factor of 10. Even if this is clearly not enough for the imaging of an Earth-like planet orbiting a Sun-like star, this is very promising as the absence of an FPM provides us with a stellar attenuation on the level of $10^{-8}$ with a very high planet throughput, on the order of $80\%$.

\begin{figure}
    \centering
   \includegraphics[width = \textwidth]{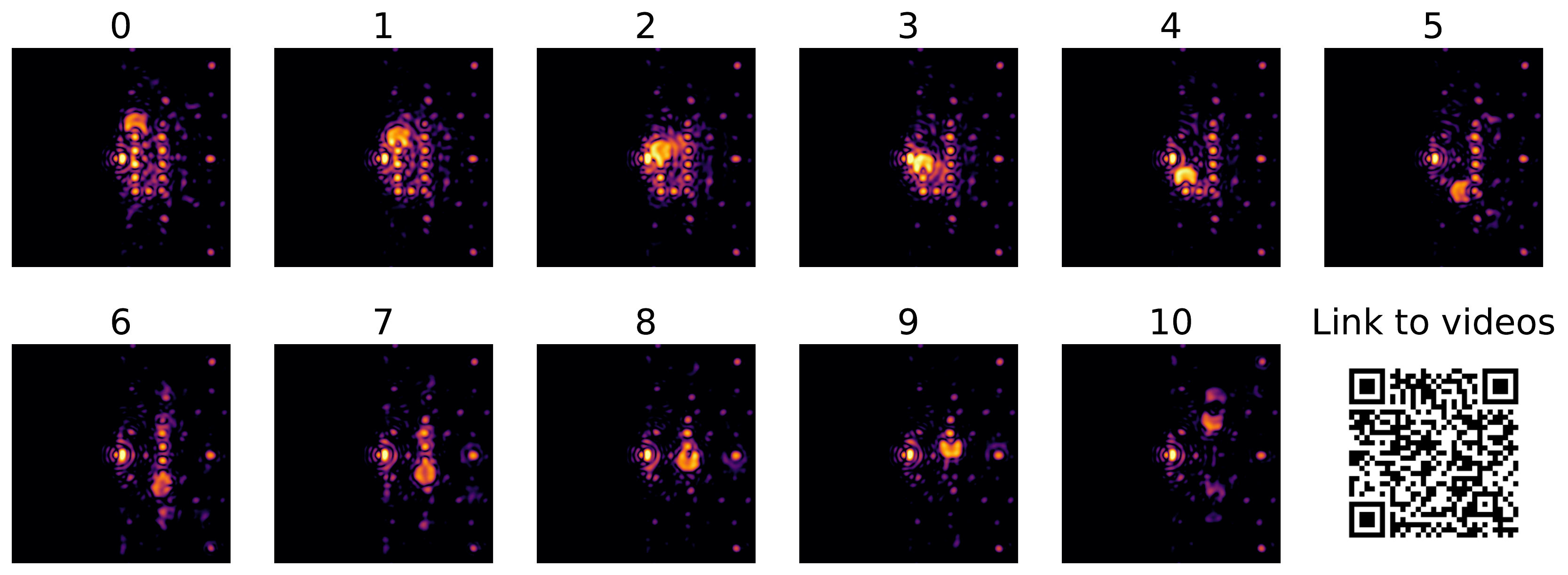}
   \caption[Image sequence] 
   {\label{fig:image_Sequence} 
    Sequence of focal-plane images of the PACMAN loop deployed on the HiCAT testbed with the PAPLC installed. An animated version of these data can be found under the associated link. In this example, PACMAN is limited to one half of the focal plane by a vertical line along the optical axis, due to the knife-edge FPM.}
   \end{figure}

\begin{figure}
   \centering
    \includegraphics[width=\linewidth]{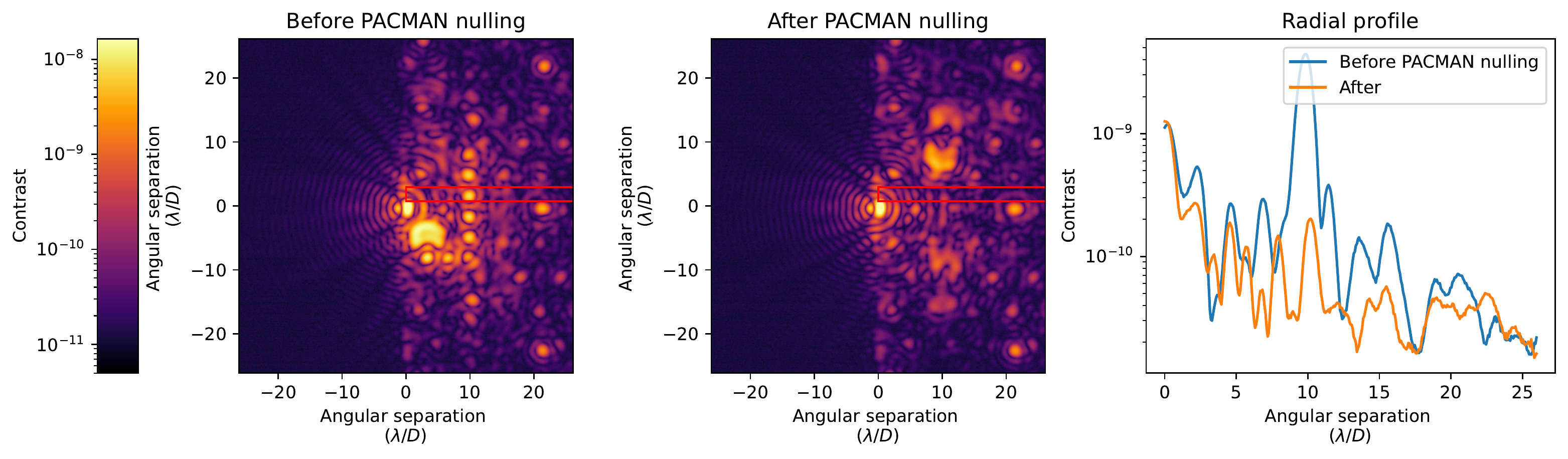}
    \caption{Experimental PACMAN run with a PAPLC coronagraph. \textit{Left}: First iteration of PACMAN, no speckles have been suppressed so far. \textit{Middle}: After a run of PACMAN. \textit{Right}: Plot of the red rectangle averaged over the vertical dimension. In this experiment, the biggest gain is observed at 10 $\lambda/D$.}
    \label{fig:paplc_cut}
    \end{figure}

However, to obtain a deeper contrast, necessary for the goals of HabWorlds for example, using an FPM is very likely mandatory. Using the current setup of HiCAT, we deployed PACMAN with the PAPLC coronagraph. An image sequence taken during this PACMAN run is shown in Fig.~\ref{fig:image_Sequence}.
The results with this configuration are shown in Fig.~\ref{fig:paplc_cut}, with images showing the speckle field before and after the release of PACMAN in the focal plane, as well as a cut to emphasize the improvement around one single speckle.
Here again, the improvement thanks to PACMAN is obvious, yielding contrast levels below $10^{-10}$, which even exceeds the theoretical performance of the PAPLC\cite{Por2020PhaseApodizedPupil}. In particular, the radial cut shows an improvement at the level of a single speckle of between one and two orders of magnitude.

\begin{figure}
    \centering
    \includegraphics[width=0.6\linewidth]{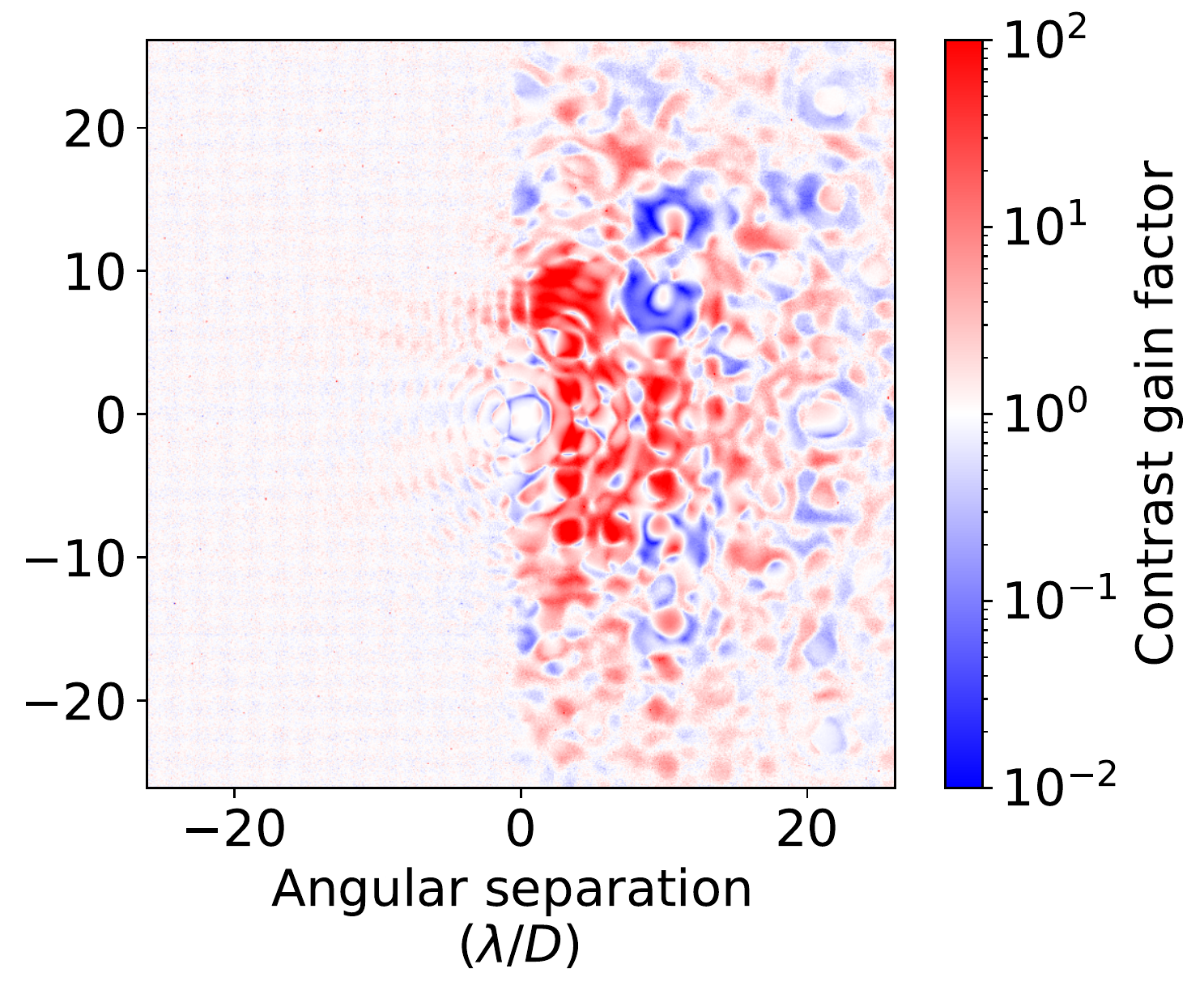}
    \caption{Experimental PAPLC contrast gain map, obtained by dividing the data from the first iteration by the data of the last iteration. Red pixels indicate an improvement, blue ones a degradation. Contrary to most other WFS\&C techniques, PACMAN is most efficient at small inner working angles (IWA), which puts it at a distinct advantage.}
    \label{fig:paplc_gain_map}
    \end{figure}

With these deep contrasts, the ghosts appearing in the focal plane become one of the main perturbation source. They are clearly visible in the image after the applied nulling. Solutions to mitigate these in a reliable manner are under study and are beyond the scope of this paper. Nevertheless, the overall gain is undeniable, as demonstrated in Fig.~\ref{fig:paplc_gain_map}.
It shows the gain in terms of contrast factor between the first and the last image of the experimental PAPLC run. Here, we obtain a substantial improvement of two orders of magnitude close to the on-axis star, at separations shorter than 10$\lambda/D$, where we expect to search for exoplanets. A more moderate gain of only one order of magnitude is also visible across the focal plane, beyond the walls of the maze, even if some degradation is also visible here and there, probably due to ghosts limiting the intervention of PACMAN.

When comparing the simulated data from Fig.~\ref{fig:sim_paplc_cut} to experimental data from Fig.~\ref{fig:paplc_cut}, we can tell they show very good agreement. We note the similar contrast levels and similar perturbations, even if the simulation tends to underestimate the number of ghosts. This is promising for future tests, including for the CLC or apodized pupil Lyot coronagraphs (APLC\cite{N'Diaye2016APODIZEDPUPILLYOT}) that will be tested on HiCAT in the near future.

\section{Discussion: Feasibility of PACMAN for future space-based missions}
\label{sec:real-data}

The real question we pose in this paper is: can PACMAN be flown on HabWorlds\footnote{``HWO'' is quite a mouthful.} to be used to image Earth-like exoplanets?

We can identify several promising avenues for improvement of PACMAN. As it learns to find its path through the speckle maze over different patterns, advancing through levels will make its speed in the focal plane increase over time. Eventually, it would reach correction frequencies that are fast enough to use PACMAN as a focal-plane correction method that stabilizes the contrast in the DH over time. We thus consider the stability problem of large, segmented space telescopes as solved. No further research is needed \footnote{\url{https://xkcd.com/2268/}}.

As shown with the data in Sec.~\ref{sec:results}, there are still some issues with the ghost evasion. Especially when moving on to broadband demonstrations, it will be crucial to investigate improved regularization methods for the Jacobian inversion in order to minimize their influence on the path-finding equations. One of the biggest potential for ghost evasion lies in the investigation of teleportation tunnels. While we did not consider them in this paper at all, their implementation would allow PACMAN to move quickly across the maze, which increases its chance to run into bonus pellets (i.e., planets). Whether such teleportation tunnels are best implemented along contours of equal spatial frequency, or along specified azimuthal angles, is out of scope of this paper.

While we have shown that PACMAN can provide us with incredible contrast levels on the order of $10^{-10}$, there are some serious obstacles to address when extrapolating its efficiency to future, space-based HCI missions. The main one is its very limited duty cycle. During the experiments presented in this paper, we measured a duty cycle of $1/365 \pm 7$ days, centered around a mean value of 1 April. Using a proprietary statistical modelling tool, we found that it tends to periodically drop to a duty cycle of $1/366 \pm 7$ days, with a period of four years, and with the same mean value. This means that it is utterly unreliable as of now and more development would have to be invested to make it a feasible candidate for focal-plane WFS\&C.

Further, with its original release in 1980, there are some doubts around the knowledge transfer to generations that will be involved in the commissioning and operation of a 2040s space telescope. While PACMAN remains very popular among millenials, especially early millenials, the authors fear that one sad day in the future, it might be considered as uncool as any reflection, activity or comment that today is acquitted with a nonchalant ``ok boomer''. As a result, the authors strongly suggest an accelerated development of the HabWorlds mission, ideally to meet a launchdate in the early 2030's.  

On a more technical side, the current technology readiness level (TLR) of PACMAN is very low. While there was indeed a ``Gimme Space'' level in the edition of Pac-Man World in the past, the system-level TRL of arcade cabinets can generally assumed to be -2. Hence, we do not see this algorithm take off to the stars any time soon unless substantial efforts are made now. Reaching a TRL level of at least 8 in the next decade should be a priority. If cubesats might be limited for arcade hardware, the imminent commercial availability Artemis launchers should make this goal relatively easy. 

\begin{figure}
    \centering
    \includegraphics[width=0.6\linewidth]{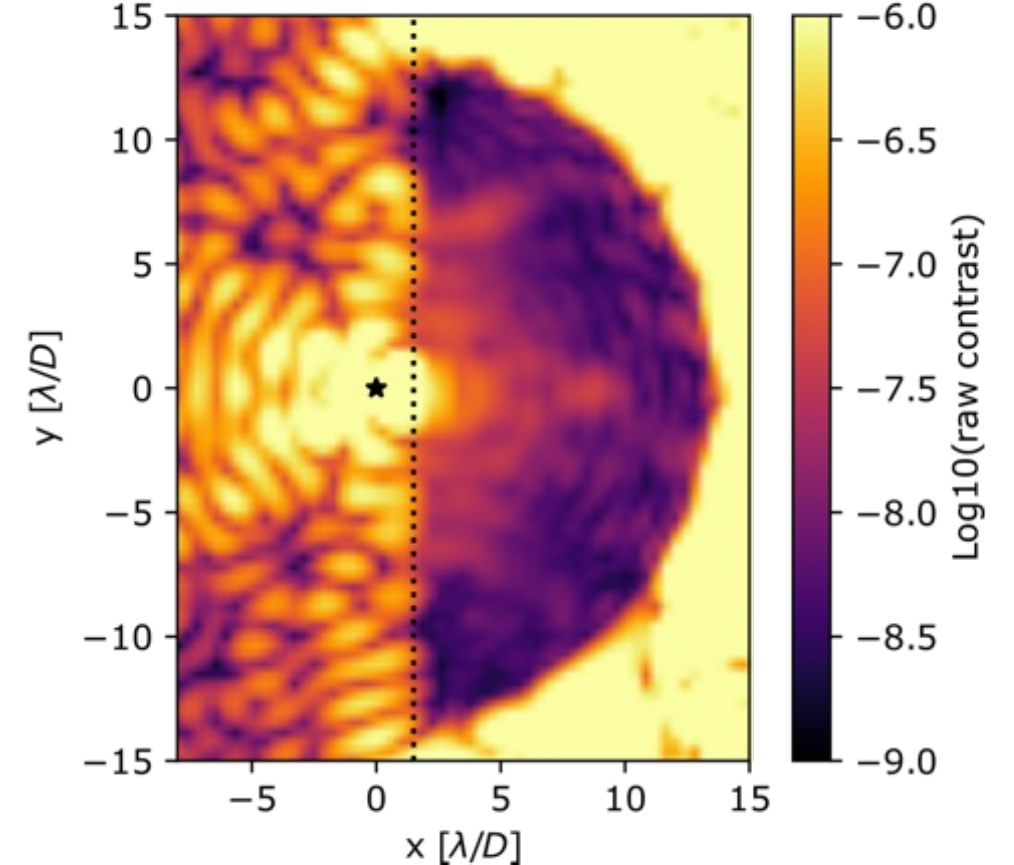}
    \caption{Best PAPLC result on HiCAT in monochromatic light to date. The PAPLC provides an IWA of 2.3 $\lambda/D$, and produces a half-DH with an average contrast of $2\times10^{-8}$ from 2-13 $\lambda/D$, and $8\times10^{-9}$ from 5-13 $\lambda/D$.\cite{Soummer2022HighContrastImager}}
    \label{fig:best_paplc_mono}
    \end{figure}

The good new is that there are more robust methods for focal-plane WFS\&C under development on HiCAT and other testbeds around the world\cite{Laginja2022CAOTIC,Mazoyer2019HighContrastTestbedsFuture}. Using the above quoted PAPLC, which provides an inner working angle (IWA) of 2.3 $\lambda/D$, HiCAT produces a half-DH with an average contrast of $2\times10^{-8}$ from 2-13 $\lambda/D$, and $8\times10^{-9}$ from 5-13 $\lambda/D$ in monochromatic light, as seen in Fig.~\ref{fig:best_paplc_mono}\cite{Soummer2022HighContrastImager}.

\section{Summary}
\label{sec:summary}

Finding Earth-like exoplanets is a hard task as it is, and it does not get any easier if yo do not allow yourself some breaks while contributing to this amazing goal. In this paper, we advocate for more fun and games in our day-to-day lives. Go for a walk in nature, read a fiction book, meet up with your friends and family. *Somehow* we have weathered a pandemic none of us has seen in their life time, and if you are still around to read these lines, you can pat yourself on the shoulder, because you made it. 

Also, we are excited for HabWorlds.

\acknowledgments 
The authors are very thankful to the extended HiCAT team (over 50 people) who have worked over the past decade to develop this testbed, through a combination of NASA and STScI-internal funding. I.L. acknowledges the support by a postdoctoral grant issued by the Centre National d'Études Spatiales (CNES) in France.

\bibliography{references}
\bibliographystyle{spiebib}

\end{document}